\documentclass[10pt,twocolumn]{IEEEtran}
\usepackage{graphicx}          
\usepackage{amsmath}
\usepackage{tabularx}
\usepackage{amsfonts}
\usepackage{cite}
\usepackage{url}
\usepackage{verbatim}
\usepackage{booktabs}
\usepackage{times,epsfig}
\usepackage{makecell}
\usepackage{graphicx}
\usepackage[normalem]{ulem}
\usepackage{multirow}
\usepackage{graphicx}
\usepackage[normalem]{ulem}
\useunder{\uline}{\ul}{}
\usepackage{psfrag}
\usepackage{subfigure}
\usepackage{stfloats}
\usepackage{setspace}
\usepackage{amssymb}
\usepackage{multirow}
\usepackage{colortbl}
\usepackage[justification=centering]{caption}
\usepackage{fancyhdr}
\usepackage[table,xcdraw]{xcolor}

\begin{document}
	
	\pagenumbering{arabic}
	
	%\title{A Secure and Reliable Transfer Learning Framework for Emerging 6G Services and	Applications}
	% Quantum Key Distribution Service Provision
	\title{Generative AI-enabled Quantum Computing Networks and Intelligent Resource Allocation}
	\author{Minrui Xu, Dusit Niyato, Jiawen Kang, Zehui Xiong, Yuan Cao, Yulan Gao, Chao Ren, and Han Yu
 \thanks{Minrui~Xu, Dusit~Niyato, Yulan Gao, and Han Yu are with the School of Computer Science and Engineering, Nanyang Technological University, Singapore 639798, Singapore (e-mail: minrui001@e.ntu.edu.sg, dniyato@ntu.edu.sg, yulan.gao@ntu.edu.sg, han.yu@ntu.edu.sg). Jiawen~Kang is with the School of Automation, Guangdong University of Technology, China (e-mail: kjwx886@163.com). Zehui~Xiong is with the Pillar of Information Systems Technology and Design, Singapore University of Technology and Design, Singapore 487372, Singapore (e-mail: zehui\_xiong@sutd.edu.sg). Chao Ren is with the School of Electrical and Electronic Engineering, Nanyang Technological University, Singapore 639798, Singapore (e-mail: chao.ren@ntu.edu.sg). Yuan Cao is with the School of Communications and Information Engineering, Nanjing University of Posts and Telecommunications, Nanjing 210003, China (e-mail: yuancao@njupt.edu.cnz).}}
	\maketitle
	\pagestyle{headings}

	\begin{abstract}
 Quantum computing networks enable scalable collaboration and secure information exchange among multiple classical and quantum computing nodes while executing large-scale generative AI computation tasks and advanced quantum algorithms. Quantum computing networks overcome limitations such as the number of qubits and coherence time of entangled pairs and offer advantages for generative AI infrastructure, including enhanced noise reduction through distributed processing and improved scalability by connecting multiple quantum devices. However, efficient resource allocation in quantum computing networks is a critical challenge due to factors including qubit variability and network complexity. In this article, we propose an intelligent resource allocation framework for quantum computing networks to improve network scalability with minimized resource costs. To achieve scalability in quantum computing networks, we formulate the resource allocation problem as stochastic programming, accounting for the uncertain fidelities of qubits and entangled pairs. Furthermore, we introduce state-of-the-art reinforcement learning (RL) algorithms, from generative learning to quantum machine learning for optimal quantum resource allocation to resolve the proposed stochastic resource allocation problem efficiently. Finally, we optimize the resource allocation in heterogeneous quantum computing networks supporting quantum generative learning applications and propose a multi-agent RL-based algorithm to learn the optimal resource allocation policies without prior knowledge.
	\end{abstract}

	
	%%%%%%%%%%%%%%%%%%%%%%%%%%%%%%%%%%%%
\section{Introduction}

% 大背景: quantum computing networks

Quantum computing networks can integrate existing networking and computing infrastructure with quantum communication and computing devices to enhance network security, scalability, and sustainability~\cite{caleffi2022distributed}. Specifically, incorporating Quantum Key Distribution (QKD) into data center networks~\cite{cao2022evolution}, especially for generative AI systems, shows significant promise in bolstering privacy protection during the training and inference phases of large foundation models~\cite{ren2023towards}. Furthermore, the interconnection of quantum devices in distributed quantum computing architectures enables the tackling of complex computational tasks, which are currently infeasible for classical computers to solve in a reasonable period of time. However, quantum computing networks are still in a developing stage. Currently, quantum channels can generate several kilobytes of secret keys per second, and Noisy Intermediate-Scale Quantum (NISQ) devices are limited to maintaining only hundreds of qubits.
% The first sentence emphasizes the potential of quantum communication in network security and quantum computing in network optimization. Specifically, the second sentence introduces QDK in data center networks for distributed machine learning systems. The third sentence highlights the potential of quantum computing in tackling computation tasks that are practically impossible for classical computers to solve in a reasonable timeframe. However, the fourth sentence discusses the current development stage of quantum communication and computing is too early for realizing fully secured, scalable, and sustainable quantum ecosystem.

In quantum computing networks, devising effective resource allocation schemes is a critical but unresolved issue, which is essential for the seamless integration of classical computing and networking infrastructure for generative AI with emerging quantum devices~\cite{chen2023simqn}. The inherent dynamics of quantum resources, such as qubit instability and decoherence, pose challenges in providing quantum cryptography and quantum computing services, which are crucial for securing and enhancing classical infrastructure. Specific challenges in quantum computing networks include uncertain fidelities of qubits and entangled pairs, the coexistence of multiple communication protocols, and the heterogeneity of networks comprising diverse nodes and connections. Addressing these challenges necessitates the efficient allocation of quantum resources. This includes allocating additional qubits for error correction and utilizing entangled pairs for quantum state purification. Furthermore, developing translation policies is vital for interoperability among networks utilizing different protocols. Effective resource allocation frameworks are also crucial to ensure seamless connectivity and functionality across various types of quantum nodes and links.
% In quantum computing networks, several challenges for efficient resource allocation are raised as there is an urgent need for resource allocation schemes in networks with quantum interconnects and quantum nodes with limited capacity. The complexity of the resource allocation problem in quantum computing networks is compounded by the simultaneous consideration of classical resources and dynamic quantum resources. Additional challenges arise from the uncertain fidelity of qubits and entangled pairs, the existence of multiple protocols in quantum networks, and the heterogeneity of quantum networks with different nodes and connections. These challenges require efficient allocation of quantum resources, including the allocation of additional qubits for error correction and entangled pairs for cleaning. In addition, translation policies are essential for bridging networks with different protocols, and resource allocation frameworks are crucial to ensure seamless coverage across different types of quantum nodes and links.
% 

Fortunately, the inherent uncertainties in quantum computing networks can be effectively modeled and addressed by utilizing reinforcement learning (RL), which is capable of modeling the dynamics of quantum links and devices~\cite{cao2020multi}. RL agents learn optimal quantum resource allocation policies through interactive engagement with the quantum computing network environment. To optimize this interaction, state-of-the-art RL algorithms employ deep, quantum, or generative neural networks for policy representation. Specifically, deep learning enables RL agents to simplify the approximation of the Q-function in extensive state and action spaces. Additionally, generative AI enables RL agents to create interaction trajectories and support long-term decision-making in quantum computing networks. Moreover, RL agents using quantum policies, parameterized by quantum neural networks, can expedite action selection in centralized quantum resource allocation scenarios. The synergistic benefits of integrating generative AI with quantum computing are manifold:
\begin{itemize}
    \item \textbf{Generative AI for quantum computing networks:} Generative AI significantly enhances the sample efficiency and training stability of intelligent resource allocation algorithms, which can be achieved through long-horizon planning and the synthesis of trajectory experiences~\cite{zhu2023diffusion}.
    \item \textbf{Quantum computing networks for generative AI:} By utilizing quantum computing and communication, the training and inference processes of generative AI models can be made more secure, rapid, and energy-efficient. Moreover, by leveraging the principles of quantum mechanics, generative AI has the potential to introduce innovative architectures and models for a wide range of general or specific tasks.
\end{itemize}
Therefore, the synergy of generative AI and quantum computing networks offers the potential to develop novel algorithms for intricate resource allocation problems, as well as expand the range of applications within quantum computing networks.
% 小背景: resource allocation - quantum channels, qubits, entanglement pairs
\begin{figure*}[t]
\vspace{-0.5cm}
    \centering
    \includegraphics[width=0.8\linewidth]{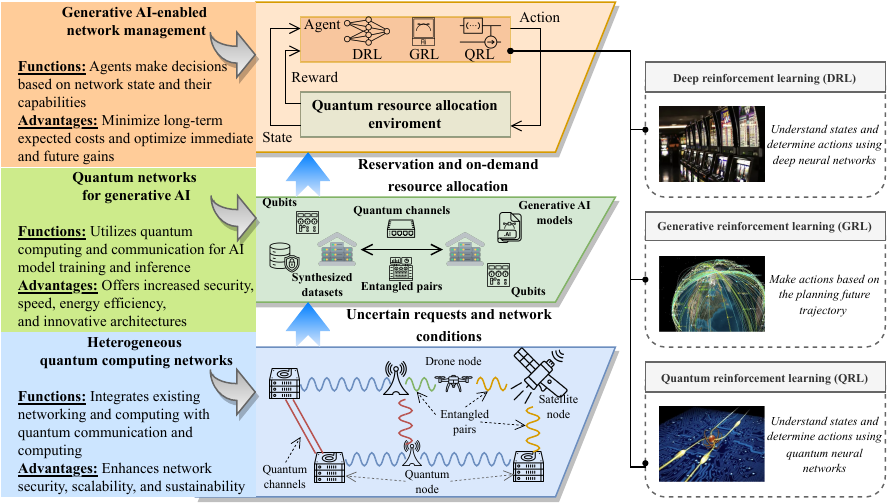}
    \caption{An overview of intelligent resource allocation in quantum computing networks.}
    \label{fig:overview}
\end{figure*}
In this article, we propose an intelligent resource allocation framework for quantum computing networks for allocating quantum resources with uncertain fidelities. First, we introduce the quantum computing networks which consist of two main services, i.e., quantum key distribution and distributed quantum computing, and three types of quantum resources, i.e., quantum channels, computation qubits, and entangled pairs. Furthermore, we propose an intelligent allocation framework for quantum resources in quantum computing networks. We introduce existing state-of-the-art RL algorithms, including, deep reinforcement learning (DRL), deep generative learning, and quantum reinforcement learning (QRL). In addition, the challenges in the intelligent resource allocation framework are highlighted. Finally, we leverage a heterogeneous quantum computing network that supports quantum generative learning applications to validate the effectiveness of the proposed intelligent resource allocation framework.

Our main contributions can be summarized as:
\begin{itemize}
    \item We propose an intelligent quantum resource allocation for quantum computing networks enabled by generative AI. On the one hand, the security and scalability of training and inference of distributed machine learning systems for generative AI can be enhanced in quantum computing networks. On the other hand, generative AI can enable more effective and robust intelligent resource allocation algorithms in quantum computing networks.
    \item We highlight the potential applications and challenges for resource allocation in quantum computing networks, where the resource allocation problem is formulated as stochastic optimization under the uncertain fidelities of qubits and entangled pairs.
    \item We present a use case of multi-agent reinforcement learning for quantum resource allocation in heterogeneous quantum computing networks to illustrate the effectiveness of the proposed framework. The experimental results demonstrate that the proposed framework can reduce the computation latency and energy consumption in providing quantum generative learning applications.
\end{itemize}
\begin{table*}[t]
\centering
\caption{Comparative Summary of Quantum Computing Network Types}
\label{tab:networks}
\resizebox{\textwidth}{!}{%
\begin{tabular}{m{.1\textwidth}<{\centering}|m{.3\textwidth}<{\raggedright}|m{.3\textwidth}<{\raggedright}|m{.3\textwidth}<{\raggedright}}
\toprule
\textbf{Network Types} & \textbf{Point-to-point Quantum Networks} & \textbf{Multi-protocol Quantum Networks} & \textbf{Heterogeneous Quantum Networks} \\
\midrule
\textbf{Objectives} & Establish secure connections between two quantum computing nodes using quantum channels & Integrate multiple experimental quantum networks into a global quantum network & Extend range and coverage, enhance network robustness \\ \hline
\textbf{Resource Types} & \begin{itemize}
    \item Quantum channels
\end{itemize} & \begin{itemize}
    \item Quantum channels
    \item Computation qubits
    \item Entangled pairs 
\end{itemize} & 
\begin{itemize}
    \item Quantum channels
    \item Entangled pairs 
\end{itemize}
\\ \hline
\textbf{Limitations} & Fidelity degradation of quantum signals over long distances & Additional protocol translation costs & High complexity in coordination and management \\
\bottomrule
\end{tabular}%
}
\end{table*}
% To address these issues: intelligent resource allocation, reinforcement learning

\section{Quantum Computing Networks}

Quantum computing networks facilitate collaboration and information exchange between multiple classical computers and quantum devices over optical backbone networks and space-air-ground integrated networks (SAGINs), as illustrated in Fig.~\ref{fig:overview}. Specifically, quantum computing networks enable the secure collaborative execution of large-scale computation tasks and advanced quantum algorithms~\cite{caleffi2022distributed}. In quantum computing networks, tasks like training and inference of foundation models can be securely conducted across multiple data centers using quantum cryptography, such as QKD which ensures secure communication by using quantum mechanics principles for key generation and distribution.  Furthermore, advanced quantum algorithms can be decomposed and compiled into smaller, manageable quantum circuits suitable for execution on NISQ devices. These NISQ devices execute quantum circuits using qubits, the fundamental units of quantum information, and quantum gates, which facilitate quantum operations.  To enable information exchange in a quantum network, entangled pairs, i.e., pairs of qubits in a quantum-mechanically linked state, are generated and distributed. This distribution allows for information exchange through a process known as entanglement swapping, a key technique for quantum communication~\cite{li2023entanglement}. By establishing hybrid connections that bridge classical and quantum devices via quantum channels, quantum computing networks can address and mitigate limitations inherent in both classical and quantum systems, including security vulnerabilities, quantum noise, and scalability issues. However, due to current hardware constraints in quantum technology, quantum circuits must be effectively mapped to correspond with the original circuit design of the advanced quantum algorithms, ensuring compatibility and optimal functionality.

\subsection{Quantum Key Distribution}

QKD is used in quantum computing networks for secure peer-to-peer communication between computation nodes~\cite{cao2022evolution}. By leveraging the non-cloning theorem which states that an unknown quantum state cannot be copied exactly in quantum mechanics, QKD focuses on cryptographic key exchange via quantum channels over fiber optic links for ground-based communications and free-space links for aerial and satellite communications.  During QKD-secured exchanges between two computational nodes, both the sender and receiver deploy their quantum communication infrastructure, establishing a hybrid link composed of classical and quantum channels. The process begins with the sender preparing and transmitting a sequence of quantum states, typically encoded in photons, through the quantum channel to the receiver. Upon receipt, the receiver performs quantum measurements on these states using specific measurement bases that are compatible with the sender's preparation method. Afterward, the sender and receiver engage in a public comparison of a subset of their measurement results over a classical channel to estimate the quantum bit error rate in the transmission. They then apply quantum error correction protocols to rectify any discrepancies in the key, followed by a process called privacy amplification. This step involves distilling a shorter, but more secure, final key from the error-corrected key, eliminating any partial information that an eavesdropper might have gained. The distinct advantage of QKD lies in its ability to generate secret keys over a distance in an information-theoretically secure manner, relying on the fundamental laws of quantum physics rather than the assumed difficulty of computational problems.
\subsection{Distributed Quantum Computing}

Distributed quantum computing involves the use of multiple interconnected quantum processors to collaborate to execute quantum circuits of advanced quantum algorithms~\cite{caleffi2022distributed}. This approach is designed to surpass the inherent limitations of single quantum processors, particularly in terms of the number of qubits available and overall computational power. The essence of distributed quantum computing lies in the coordination and execution of quantum algorithms across a network of interconnected processors. In this setup, each quantum processor is assigned a specific portion of the overall computational task. These processors then operate simultaneously, executing their respective components and communicating through a coherent quantum network. Such parallel and coordinated execution harnesses the collective computational power of the network, effectively amplifying the capabilities beyond what individual processors can achieve.
Resource requirements of distributed quantum computing can be estimated during the compiling of quantum circuits and realized during the execution through quantum Message Passing Interface (MPI) via entanglement swapping. Furthermore, the integration of distributed quantum computing into existing workflow models is facilitated by extending workflow languages. This extension enables the mapping of complex quantum computing tasks to native modeling constructs, ensuring both the portability of these workflows across different quantum computing platforms and ease of orchestration and execution. % These advancements in distributed quantum computing are crucial in realizing the full potential of quantum computing, especially in addressing computational problems that are impossible for classical computers.

\subsection{Quantum Resources}

As listed in TABLE~\ref{tab:networks}, there are three main types of quantum resources in quantum computing networks, including quantum channels, computation qubits, and entangled pairs.
% \usepackage{graphicx}
% \begin{table*}[]
% \centering
% \caption{Objectives, resource types, and limitations of different types of quantum computing networks.}
% \label{tab:networks}
% \resizebox{\textwidth}{!}{%
% \begin{tabular}{|m{.1\textwidth}<{\centering}|m{.3\textwidth}<{\raggedright}|m{.3\textwidth}<{\raggedright}|m{.3\textwidth}<{\raggedright}|}
% \hline
% Network types & Point-to-point quantum networks & Multi-protocol quantum networks & Heterogeneous quantum networks \\ \hline
% Objectives & Establish secure connections between two quantum computing nodes using   quantum channels & Integrated multiple experimental quantum networks to a global   quantum network & Extend range and coverage while Enhance network robustness \\ \hline
% Resource types & Quantum channels & Quantum channels, computation qubits, entangled pairs & Quantum channels and entangled pairs \\ \hline
% Limitations & Fidelity degradation of quantum signals over long distances & Extra protocol translation cost & High complexity in coordination and management \\ \hline
% \end{tabular}%
% }
% \end{table*}

\subsubsection{Quantum Channels}

To establish secure communication links between two computing nodes, quantum channels can be utilized to exchange secure keys for message transmission. Quantum channels facilitate the distribution of symmetric keys encoded as quantum states, enhancing communication security against eavesdropping attacks and quantum attacks.  To generate quantum channels over optical transport networks, a suite of specialized components is required, including QKD transmitters, QKD receivers, trusted relays, untrusted relays, local key managers, and the security infrastructure~\cite{cao2022evolution}. Specifically, QKD transmitters prepare and send quantum states while QKD receivers detect these states. Trusted relays are secure nodes that assist in extending the range of quantum communication, while untrusted relays facilitate quantum repeater technology without the need for full trust in the relay. 
% Additionally, a local key manager oversees the generation and management of encryption keys, and a comprehensive security infrastructure ensures the overall protection of the quantum communication network. Terrestrial quantum channels are built upon optical backbone networks which traditionally developed for classical data transmission are being adapted to support quantum communication by transmitting quantum states alongside classical data. 
As more quantum channels are allocated within these networks, there is a corresponding increase in the secure key rates, which is crucial for effective quantum communication. However, a significant challenge arises from the inherent point-to-point nature of quantum communication. Unlike classical channels that can broadcast to multiple recipients, quantum channels typically connect just two nodes directly. To address the increasing demands, it is crucial to allocate and reserve quantum channel resources according to user requests flexibly to improve the scalability of quantum computing networks.
% These components are responsible for generating, transmitting, and receiving quantum data. Currently, quantum channel resources can be allocated from Optical Backbone Networks. As the number of allocated quantum channels increases, secure key rates also increase. However, due to the point-to-point nature of quantum communication, quantum channel resources need to be reserved and allocated on-demand based on user requirements to improve the scalability of quantum computing networks.

\subsubsection{Computation Qubits}

In each quantum processor, a certain number of qubits are generated to encode and process classical information and perform operations using quantum gates~\cite{mao2023qubit}. While qubits can be used to run quantum algorithms for solving complex computational problems more efficiently than classical computers, the limited quantity of physical qubits, errors, and decoherence pose challenges in scaling quantum computing to achieve quantum advantages. Consequently, qubit allocation becomes a critical issue in quantum computing networks, where logical qubits, which represent the abstract quantum information, can be mapped onto physical qubits. This process is referred to as qubit allocation. Furthermore, the qubit allocation problem can be conceptualized as analogous to the quadratic assignment problem in classical optimization theory, which is known to be NP-hard, and is computationally challenging to solve as the size of the input increases. The primary objective of qubit allocation is to find an optimal mapping strategy, which aims to minimize the impact of quantum noise and errors while simultaneously maximizing the efficiency and scalability of the quantum computing network.

\subsubsection{Entangled Pairs}

The exchange of information between two quantum nodes can be achieved through entanglement swapping of entangled pairs~\cite{li2022fidelity}, which are pairs of qubits that are quantum-mechanically linked, such that the state of one (regardless of the distance separating them) instantaneously influences the state of the other. When entanglement connections span long distances, the fidelity of entangled pairs may degrade due to imperfect measurements on quantum repeaters during entanglement swapping operations. To ensure the fidelity of entangled pairs, the iterative purification operation is applied to the entangled pairs. This purification process involves the use of multiple entangled pairs to distill a smaller number of pairs with higher fidelity. The fidelity of a purified entangled pair is quantitatively assessed by comparing its original fidelity with the fidelity post-purification. By iteratively applying this purification process, the fidelity of entanglement connections can be enhanced to meet the requisite quality standards for reliable quantum communication. The purification process has its trade-off between generating additional entangled pairs and those used for the actual information exchange. 

\subsection{Potential Applications}

\subsubsection{Distributed Machine Learning Systems}

Distributed machine learning systems within quantum computing networks offer a robust framework for collaborative AI model training and inference across multiple computing nodes. During the collaborative training process across multiple computing nodes, the raw data used for training is kept in local databases for training local models~\cite{ren2023towards}. Upon completion of local model training, these computing nodes engage in a process of model sharing, which involves transmitting the trained models to peer nodes or central servers for model aggregation. In this way, the integrity of the original data can be preserved and the creation of a more robust and comprehensive global model can be trained through this aggregation process. However, the transmission of these models poses a significant risk, particularly in terms of the security of valuable and potentially sensitive AI models. To mitigate the risk of these models being intercepted, monitored, or stolen during transmission, QKD links are established between computing nodes.

\subsubsection{Sensor Networks, Smart Grids and Internet of Things}

Sensor networks, smart grids, and the Internet of Things are constantly collecting diverse types of data from environmental parameters to user interactions for real-time monitoring, analysis, and decision-making. During the process of data transmission, some highly sensitive sensing data may be at risk of being intercepted and stolen, which can have an impact on high-risk applications. For example, in secure smart grids, where data on power consumption and distribution needs to be protected, quantum computing networks can provide secure communication channels through quantum key distribution~\cite{kong2020review}. This ensures that the data remains confidential and cannot be tampered with, enhancing the overall security of the smart grid system. By harnessing the advanced computational power of quantum computing, the analysis of voluminous sensing data can be performed with heightened efficiency and precision. This not only leads to more accurate decision-making but also contributes to substantial resource savings.

\section{Generative AI-enabled Intelligent Resource Allocation for Quantum Computing Networks}

% \subsection{Intelligent Resource Allocation Schemes}

With the benefits of being adaptable to changing environments, RL is used to enhance the efficiency of resource allocation in quantum computing networks. For example, in the resource allocation problem of QKD networks with multiple tenants, RL-based algorithms are proposed to improve the blocking probability of tenant requests and the utilization of secret-key resources~\cite{cao2020multi}. Specifically, RL agents can model the dynamic environments as a Markov Decision Process (MDP), considering states, actions, rewards, and transition probabilities. In the environment of quantum computing networks, the MDP can usually be defined as follows:
\begin{itemize}
    \item \textbf{State space:} The state space in quantum computing networks encompasses key configurations and situations of the network, which includes real-time monitoring of computing nodes, communication links, application requests, network topology, the number of qubits in quantum nodes, the quantity of both quantum and classical channels in quantum links, and the operational protocols of the quantum network.
    \item \textbf{Action space:} The action space is defined by the decisions made by RL agents at each time slot, which involves the allocation of available quantum resources, such as quantum channels, computation qubits, and entangled pairs, for the efficient execution of QKD protocols and quantum algorithms.
    \item \textbf{Reward function:} The reward function is formulated to reflect the net utility derived from providing quantum communication and computing services, which depends on both the benefits of service delivery and the costs associated with provisioning quantum resources.
    \item \textbf{State transition probability:} Based on the present state of the quantum network environment, RL agents determine the allocation of quantum resources. Following this decision, the environment evolves into a subsequent state, guided by a state transition probability model. 
    % This transition encapsulates the inherent uncertainties in quantum computing networks, including fluctuations in resource demands and the challenges of on-demand resource allocation aimed at satisfying all requests.
\end{itemize}
Subsequently, RL agents learn the policy to optimize the long-term expected efficiency of resource utilization under changing environmental conditions.  A summary of existing RL algorithms can be found in TABLE~\ref{tab:algorithm}, which introduces advantages, target scenarios, complexity, and famous algorithms of different schemes.

% \usepackage{graphicx}
% \begin{table*}[]
% \centering
% \caption{The comparison of different learning-based quantum resource allocation schemes (N = number of classic computing operations, H = length of trajectories, Q = number of quantum computing operations).}
% \label{tab:algorithm}
% % \resizebox{\textwidth}{!}{%
% \small\begin{tabular}{|m{.15\textwidth}<{\centering}|m{.25\textwidth}<{\raggedright}|m{.25\textwidth}<{\raggedright}|m{.25\textwidth}<{\raggedright}|}
% \hline
% Schemes & Deep reinforcement learning & Deep generative learning & Quantum reinforcement learning \\ \hline
% Advantages & Efficient management of high-dimensional state and action spaces & Improved sample quality and training stability &  Offer quantum advantage with quantum policy representation and optimization \\ \hline
% Target scenarios & Decentralized resource allocation & Long-horizon planning, Data synthesizing & Centralized resource allocation \\ \hline
% Computation and sample complexity & \textit{O(N$^2$)} & \textit{O(HN)} & \textit{O(QN)} \\ \hline
% Algorithms & Deep Q-network, Deep Deterministic Policy Gradient & Decision transformer, Decision diffuser & QiRL, VQC-based Q-learning, DDQL \\ \hline
% \end{tabular}%
% % }
% \end{table*}

\begin{table*}[t]
\centering
\caption{Comparison of Different Learning-Based Quantum Resource Allocation Schemes (\textit{N} = number of classic computing operations, \textit{H} = length of trajectories, \textit{Q} = number of quantum computing operations)}
\label{tab:algorithm}
\resizebox{\textwidth}{!}{%
\small\begin{tabular}{m{.15\textwidth}<{\centering}|m{.25\textwidth}<{\raggedright}|m{.25\textwidth}<{\raggedright}|m{.25\textwidth}<{\raggedright}}
\toprule
\textbf{Schemes} & \textbf{Deep Reinforcement Learning} & \textbf{Deep Generative Learning} & \textbf{Quantum Reinforcement Learning} \\
\midrule
\textbf{Advantages} & Efficient management of high-dimensional state and action spaces & Improved sample quality and training stability & Offers quantum advantage with quantum policy representation and optimization \\ \hline
\textbf{Target Scenarios} & Decentralized resource allocation & Long-horizon planning via trajectory synthesizing & Centralized resource allocation \\ \hline
\textbf{Computation and Sample Complexity} & \(O(N^2)\) & \(O(HN)\) & \(O(QN)\) \\ \hline
\textbf{Algorithms} & Deep Q-network \& Deep Deterministic Policy Gradient & Decision transformer \& Decision diffuser & Quantum-inspired RL \& VQC-based Q-learning \\
\bottomrule
\end{tabular}%
}
\end{table*}

\subsubsection{Deep Reinforcement Learning}

With the expanding scale of quantum networks, both the state space and action space in the MDP formulated by RL agents enlarge correspondingly. This growth presents a challenge in learning an optimal policy due to the ``curse of dimensionality." Fortunately, using deep neural networks to represent complex policies, DRL addresses this challenge to effectively manage high-dimensional state and action spaces by integrating deep learning techniques with traditional RL. DRL's use of neural networks empowers agents to effectively manage high-dimensional state and action spaces. For instance, the Deep Q-network (DQN) algorithm, proficient in outputting discrete actions, leverages deep neural networks for approximating Q-values in complex environments. Conversely, the Deep Deterministic Policy Gradient (DDPG) algorithm, suitable for continuous action spaces, employs deep neural networks in both its actor and critic components. These capabilities significantly reduce the computational burden and time required for optimal resource allocation decisions in quantum networks.

% As the scale of quantum networks continues to grow, the state space and action space in the MDP established by RL agents also keep expanding, which makes it difficult for RL agents to learn an optimal policy due to the curse of dimensionality. Deep reinforcement learning (DRL) integrates deep learning techniques with RL to represent the policy with deep neural networks. This enables DRL agents to handle high-dimensional state space and action space using neural networks, thus requiring only limited computation and time to make appropriate resource allocation decisions.

\subsubsection{Deep Generative Learning}

With the inherent capability of generating trajectories for extended decision-making periods, generative AI plays a pivotal role in simulating prospective states in dynamic environments~\cite{zhu2023diffusion}. Specifically, diffusion models, a type of generative AI, excel in optimizing complete trajectories through iterative denoising steps. This process allows for the generation of intricate action distributions that surpass the limitations of unimodal distributions. Furthermore, these models demonstrate significant advantages such as enhanced sample quality and improved training stability. In scenarios involving long-horizon planning and data synthesis, generative models are particularly beneficial. For example, the Decision Transformer uses a Generative Pre-trained Transformer (GPT) architecture to predict actions autoregressively, based on the desired rewards, past states, and actions. Therefore, it can generate trajectories that match desired returns on RL tasks based on GPT's generative capabilities. DRL agents can extrapolate and generate higher returns than the maximum episode return available in the dataset, showcasing their ability to generate novel and optimized actions. In addition, the Decision Diffuser is a conditional generative model for sequential decision-making that simplifies the decision-making pipeline by framing it as conditional generative modeling, eliminating the need for reinforcement learning. It outperforms existing offline RL approaches on standard benchmarks by sampling for high returns~\cite{zhu2023diffusion}. This allows it to generate future actions that achieve the desired return. Generative AI data synthesizers significantly augment the quantity and variety of training data available for DRL agents. This expansion is critical during the policy improvement phase, where agents rely on diverse experiences accumulated during the exploration phase.
However, it is important to note the heightened computation and sample complexity associated with these approaches.

\subsubsection{Quantum Reinforcement Learning}

Beyond using DNNs to represent the policy of RL agents, quantum reinforcement learning utilizes quantum computing to calculate the policies represented by quantum neural networks~\cite{ren2022nft}. Leveraging fundamental quantum mechanical principles like superposition and entanglement, QNNs exhibit enhanced generalization capabilities, offering significant improvements in policy exploration and exploitation tasks. Furthermore, QRL introduces advantages such as improved sample quality and training stability, particularly beneficial in centralized quantum resource allocation for large-scale and complex networks. With these quantum advantages, notably computational acceleration, QRL policies attain faster convergence rates and superior performance, characterized by reduced inference delay. Additionally, unlike traditional RL approaches, QRL tends to exhibit lower computational and sample complexity, making it a more efficient solution for resource allocation in quantum computing networks.

\subsection{Challenges in Resource Allocation for Quantum Computing Networks}

\subsubsection{Uncertain Fidelities of Qubits and Entangled Pairs}

During the process of allocating quantum resources, e.g., qubits and entangled pairs, to execute quantum services for users in quantum computing networks, there can be challenges in efficiently provisioning resources due to fluctuations in user demand and quantum circuit requirements \cite{kaewpuang2023elastic}. These fluctuations can be attributed to factors like varying computational requirements, the specifics of quantum algorithms, and dynamic changes in quantum network topologies. Additionally, natural noise inherent in quantum computing and communication systems further complicates these fluctuations. A primary source of this uncertainty is the fidelity of qubits and entangled pairs, as it determines the reliability and accuracy of quantum states during computation and transmission. For remote collaboration over long distances between computing nodes, maintaining high-quality entanglement and sustained coherence times is essential for the successful completion of distributed quantum computing tasks. To ensure the required fidelities, strategies such as allocating additional qubits for quantum error correction techniques and employing entanglement purification protocols with additional entangled pairs are implemented.

\subsubsection{Multiple-Protocols Quantum Networks}

In the current developmental phase, various small-scale experimental quantum networks have been established, primarily focusing on testing and refining quantum communication and computing algorithms \cite{cao2022single}. However, the use of disparate networking protocols in these quantum networks presents a significant barrier to achieving global interconnectedness and interoperability. To facilitate the realization of large-scale, interconnected quantum networks, protocol translation policies have been proposed. These policies aim to serve as a bridge, enabling communication and coordination among small-scale quantum networks operating on different protocols. While protocol translation policies are instrumental in ensuring secure and successful interoperability between different QKD protocols, thus enhancing network scalability, they necessitate additional quantum resources. These resources are critical for optimizing the probability of successful translation and minimizing the risks associated with quantum data relaying.

\subsubsection{Heterogeneous Quantum Networks}

The heterogeneous quantum networks comprise a diverse array of quantum nodes, such as terrestrial nodes, satellite nodes, and unmanned aerial vehicle (UAV) nodes, interconnected by various quantum links, such as fiber links and free-space links \cite{xu2022quantum}. These networks not only significantly extend the coverage range of quantum communication but also enhance the practical usability and adaptability of quantum applications, such as QKD and entanglement distribution, across varied environments. However, the mobility of satellite and UAV nodes introduces greater complexity and uncertainty in resource allocation compared to more static terrestrial networks, primarily due to dynamic network topologies and variable link stability. To ensure seamless coverage and integration across space-based, aerial, and terrestrial segments, sophisticated resource allocation frameworks are essential. These frameworks should be capable of enhancing real-time data transmission, enabling flexible and adaptive satellite- and UAV-based QKD services, and employing a universal provisioning model to optimize resource utilization and minimize operational costs of the QKD services.

\section{Use Case: Multi-agent Reinforcement Learning-based Resource Allocation for Quantum Generative Learning}

% background in quantum generative learning, quantum autoencoder

\begin{figure*}
\vspace{-0.5cm}
    \centering
    \includegraphics[width=0.8\linewidth]{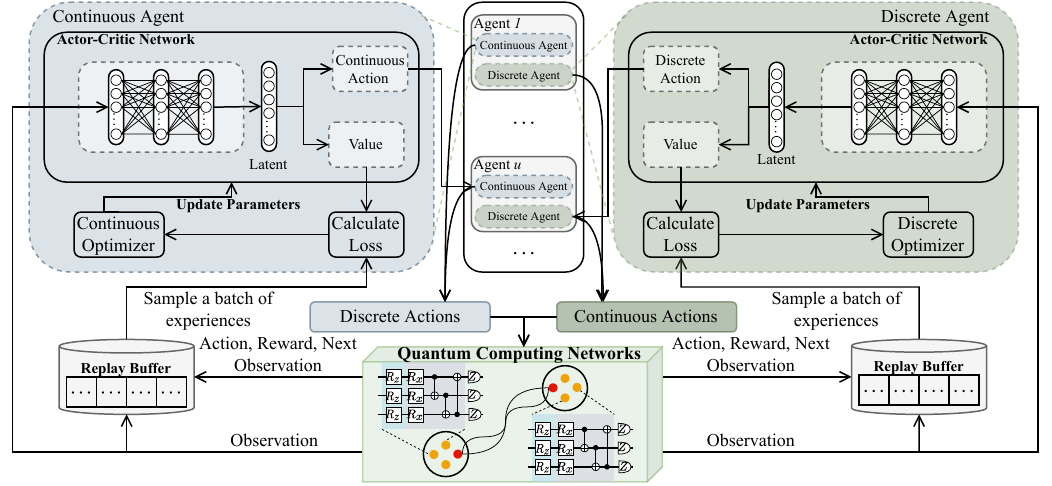}
    \caption{The intelligent resource allocation framework with discrete and continuous action spaces.}
    \label{fig:framework}
\end{figure*}

In heterogeneous quantum computing networks, comprising mobile, edge, and cloud nodes, a collaborative framework is essential to deliver a variety of quantum services. As a notable service, quantum-generative learning can generate content with quantum-generative adversarial networks and compress information through quantum autoencoders. This technique is a quantum analog of classical autoencoders, known for compressing and then reconstructing input data. In quantum generative learning, the quantum autoencoder efficiently processes complex quantum data, offering advantages in speed and dimensionality over classical counterparts. This process, however, is computationally demanding, necessitating extensive collaboration among the varied computing nodes within the quantum network. Given the diversity of these nodes in terms of computational power and functionality, allocating resources optimally poses a significant challenge. Centralized approaches to resource allocation are impractical due to the complexity and scalability issues associated with managing such a diverse and dynamic network~\cite{caleffi2022distributed}. Therefore, a decentralized approach as shown in Fig.~\ref{fig:framework}, possibly utilizing multi-agent reinforcement learning (MARL), is better suited. In this context, each agent (node) learns to make decisions about resource allocation based on the network’s current state and its own capabilities, leading to a more efficient and better scalable solution for supporting quantum generative learning applications.

% Heterogeneous quantum networks consist of mobile devices, edge nodes, and cloud nodes.
In the multi-agent scenario for resource allocation in quantum generative learning, each agent navigates a complex environment, modeled as a Partially Observable Markov Decision Process (POMDP), which is modeled for decision-making problems where the agent has incomplete knowledge about the environment's state. In this context, each agent's observations include local service statuses and computing capacities, providing limited but crucial information for decision-making. Each agent employs a policy, i.e., a strategy or set of rules, to make resource allocation decisions. These decisions involve determining whether to offload quantum generative learning tasks to either edge or cloud nodes and deciding the proportion of tasks to offload. This offloading strategy is critical as it affects the computational load distribution and overall network efficiency. The reward signal for each agent is the global resource cost, incentivizing agents to find cost-effective resource allocation strategies. The experiences from these interactions are stored in replay buffers, which are used to refine and improve the agents' policies over time. Furthermore, DRL agents learn from past experiences in replay buffers, aiding in the optimization of future decisions. The primary objective of each agent is to minimize the long-term expected cost within the quantum computing networks which involves balancing immediate rewards with future gains. By optimizing their policies, the agents aim to achieve efficient resource allocation in the dynamic environment of quantum computing networks.

\begin{figure}
    \centering
    \includegraphics[width=1\linewidth]{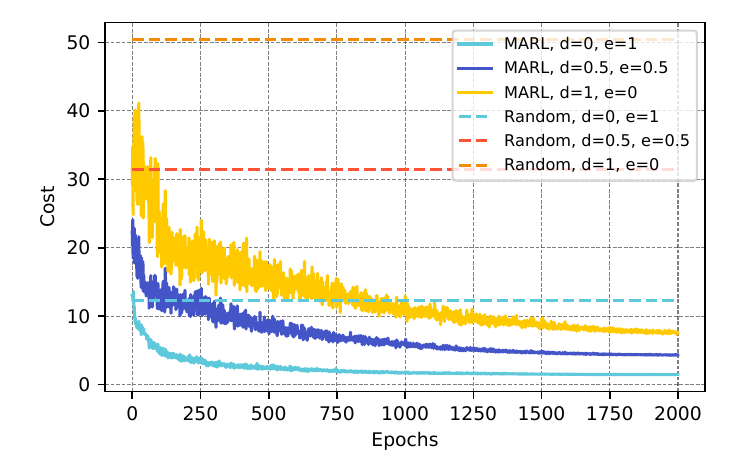}
    \caption{The convergence of proposed multi-agent RL algorithms under different QoS settings, where \textit{d} is the weight of latency and \textit{e} is the weight of energy consumption.}
    \label{fig:usecase}
\end{figure}

In quantum computing networks, each computing node can run quantum generative learning applications, such as quantum autoencoders~\cite{romero2017quantum}. Quantum autoencoders are a type of quantum circuit designed to compress quantum states onto fewer qubits while preserving the information from the initial state. In this way, the dimensionality of the information can be reduced while the fidelity is maintained. For each quantum autoencoder, there is an input layer, a bottleneck layer for compression, and an output layer for reconstruction. For instance, to compress n qubits into \textit{k} qubits using a quantum autoencoder, there are \textit{n} qubits for the input state, \textit{n-k} qubits for the reference state, and one auxiliary qubit. Therefore, the total number of qubits required would be \textit{2n-k+1} qubits. In the experiment, each computing node executes the quantum autoencoders to compress information from [6,9] qubits to [3,5] qubits. In the networks, the number of mobile nodes is set to 10, the number of edge nodes is set to 5, and there is one cloud node. The cost is calculated based on the execution latency and energy consumption according to the quality of service (QoS) settings. We leverage multi-agent reinforcement learning algorithms to allow decentralized decision-making among computing nodes. As we can observe from Fig.~\ref{fig:usecase}, the proposed MARL algorithm can converge the optimal policy for resource allocation in heterogeneous networks. Compared to the random policy, the cost can be reduced by at least 80\% under different QoS settings. Therefore, the proposed algorithm can effectively reduce the system cost by allocating quantum resources properly for running quantum generative learning applications.

% 建模出来的MDP和算法

% 实验设置和结果

% \section{Future Research Directions}

% \subsection{Simulation Environments}
% \subsection{Multi-agent Learning Algorithm}

\section{Conclusions}

In this article, we have proposed an intelligent resource allocation framework for quantum computing networks with uncertain quantum resources and service demands. We have discussed several services in quantum computing networks for improving the security and sustainability of existing networking and computing infrastructure. Furthermore, to improve the scalability of quantum computing networks, we have proposed an intelligent resource allocation framework, which can leverage RL agents, from generative learning to quantum learning, to model the environments and learn the optimal allocation policy without prior knowledge. Finally, we have designed a use case of quantum generative learning service provisioning in heterogeneous quantum computing networks and proposed a MARL algorithm to learn the optimal policy.

\bibliographystyle{ieeetr}
\bibliography{main}
\end{document}